\RequirePackage{fix-cm}

\documentclass[twocolumn,epjc3-upd,fleqn]{svjour3}
\usepackage{graphicx}
\smartqed  
\RequirePackage{graphicx}
\RequirePackage{subfigure}
\usepackage{rotating}
\usepackage{amsmath}
\usepackage{mathtools}
\RequirePackage[colorlinks,citecolor=blue,urlcolor=blue,linkcolor=blue]{hyperref}
\usepackage{doi}
\usepackage[]{units}
\usepackage{xspace}
\usepackage[switch]{lineno}
\usepackage{cite}
\usepackage{comment}
\usepackage{url}
\newcommand{\DBD}{0$\nu\beta\beta$}
\newcommand{\LMO}{Li$_{2}${}$^{100}$MoO$_4$}

\begin{document}

\title{Optimization of the first CUPID detector module}

\author{A.~Armatol\thanksref{CEA_IRFU_France}
\and
C.~Augier\thanksref{IP2I_France}
\and
F.~T.~Avignone~III\thanksref{UofSC_US}
\and
O.~Azzolini\thanksref{LNL_Italy}
\and
M.~Balata\thanksref{LNGS_Italy}
\and
K.~Ballen\thanksref{LNL_Italy}
\and
A.~S.~Barabash\thanksref{ITEP_Russia}
\and
G.~Bari\thanksref{SdB_Italy}
\and
A.~Barresi\thanksref{MIB_Italy,UniMIB_Italy}
\and
D.~Baudin\thanksref{CEA_IRFU_France}
\and
F.~Bellini\thanksref{SdR_Italy,SURome_Italy}
\and
G.~Benato\thanksref{LNGS_Italy}
\and
M.~Beretta\thanksref{UCB_US}
\and
M.~Bettelli\thanksref{CNR-IMM_Italy}
\and
M.~Biassoni\thanksref{MIB_Italy}
\and
J.~Billard\thanksref{IP2I_France}
\and
V.~Boldrini\thanksref{CNR-IMM_Italy,SdB_Italy}
\and
A.~Branca\thanksref{MIB_Italy,UniMIB_Italy}
\and
C.~Brofferio\thanksref{MIB_Italy,UniMIB_Italy}
\and
C.~Bucci\thanksref{LNGS_Italy}
\and
J.~Camilleri\thanksref{VT_US}
\and
C.~Capelli\thanksref{LBNL_US}
\and
S.~Capelli\thanksref{MIB_Italy,UniMIB_Italy}
\and
L.~Cappelli\thanksref{LNGS_Italy}
\and
L.~Cardani\thanksref{SdR_Italy}
\and
P.~Carniti\thanksref{MIB_Italy,UniMIB_Italy}
\and
N.~Casali\thanksref{SdR_Italy}
\and
E.~Celi\thanksref{LNGS_Italy,GSSI}
\and
C.~Chang\thanksref{ANL_US}
\and
D.~Chiesa\thanksref{MIB_Italy,UniMIB_Italy}
\and
M.~Clemenza\thanksref{MIB_Italy,UniMIB_Italy}
\and
I.~Colantoni\thanksref{CNR-NANOTEC,SdR_Italy}
\and
S.~Copello\thanksref{SdG_Italy,UnivGenova}
\and
E.~Craft\thanksref{Yale_US}
\and
O.~Cremonesi\thanksref{MIB_Italy}
\and
R.~J.~Creswick\thanksref{UofSC_US}
\and
A.~Cruciani\thanksref{SdR_Italy}
% ...
}

\institute{IRFU, CEA, Universit\'e Paris-Saclay, Saclay, France\label{CEA_IRFU_France}
\and
Univ Lyon, Université Lyon 1, CNRS/IN2P3, IP2I, Lyon, France\label{IP2I_France}
\and
University of South Carolina, Columbia, SC, USA\label{UofSC_US}
\and
INFN Laboratori Nazionali di Legnaro, Legnaro, Italy\label{LNL_Italy}
\and
INFN Laboratori Nazionali del Gran Sasso, Assergi (AQ), Italy\label{LNGS_Italy}
\and
National Research Centre Kurchatov Institute, Institute for Theoretical and Experimental Physics, Moscow, Russia\label{ITEP_Russia}
\and
INFN Sezione di Bologna, Bologna, Italy\label{SdB_Italy}
\and
INFN Sezione di Milano - Bicocca, Milano, Italy\label{MIB_Italy}
\and
University of Milano - Bicocca, Milano, Italy\label{UniMIB_Italy}
\and
INFN Sezione di Roma, Rome, Italy\label{SdR_Italy}
\and
Sapienza University of Rome, Rome, Italy\label{SURome_Italy}
\and
University of California, Berkeley, CA, USA\label{UCB_US}
\and
CNR-Institute for Microelectronics and Microsystems, Bologna, Italy\label{CNR-IMM_Italy}
\and
Virginia Polytechnic Institute and State University, Blacksburg, VA, USA\label{VT_US}
\and
Lawrence Berkeley National Laboratory, Berkeley, CA, USA\label{LBNL_US}
\and
Gran Sasso Science Institute, L'Aquila, Italy\label{GSSI}
\and
Argonne National Laboratory, Argonne, IL, USA\label{ANL_US}
\and
CNR-Institute of Nanotechnology, Rome, Italy\label{CNR-NANOTEC}
\and
INFN Sezione di Genova, Genova, Italy\label{SdG_Italy}
\and
University of Genova, Genova, Italy\label{UnivGenova}
\and
Yale University, New Haven, CT, USA\label{Yale_US}
\and
INFN Laboratori Nazionali di Frascati, Frascati, Italy\label{LNF_Italy}
% ...
\and
Universit\'e Paris-Saclay, CNRS/IN2P3, IJCLab, Orsay, France\label{IJCLab_France}
\and
Northwestern University, Evanston, IL, USA\label{NWU_US}
\and
Massachusetts Institute of Technology, Cambridge, MA, USA\label{MIT_US}
% ...
\and
Boston University, Boston, MA, USA\label{BU_US}
\and
California Polytechnic State University, San Luis Obispo, CA, USA\label{CalPoly_US}
% ...
\and
University of California, Los Angeles, CA, USA\label{UCLA_US}
\and
Drexel University, Philadelphia, PA, USA\label{Drexel_US}
\and
Johns Hopkins University, Baltimore, MD, USA\label{JHU_US}
% ...
\and
Nikolaev Institute of Inorganic Chemistry, Novosibirsk, Russia\label{NIIC_Russia}
\and
INFN Sezione di Padova, Padova, Italy\label{PD_Italy}
\and
Univ. Grenoble Alpes, CNRS, Grenoble INP, SIMAP, Grenoble, France\label{SIMaP_Grenoble_France}
\and
University of Bologna, Bologna, Italy\label{UnivBologna_Italy}
}

\thankstext{e1}{Previously at: IRFU, CEA, Universit\'e Paris-Saclay, Saclay, France}
\thankstext{e2}{Presently at: IRFU, CEA, Universit\'e Paris-Saclay, Saclay, France}

\date{Received: date / Accepted: date}
\maketitle

\begin{abstract}
CUPID will be a next generation experiment searching for the neutrinoless double $\beta$ decay, whose discovery would establish the Majorana nature of the neutrino. Based on the experience achieved with the CUORE experiment, presently taking data at LNGS, CUPID aims to reach a background free environment by means of scintillating \LMO \ crystals coupled to light detectors. Indeed, the simultaneous heat and light detection allows us to reject the dominant background of $\alpha$ particles, as proven by the CUPID-0 and CUPID-Mo demonstrators.\\
In this work we present the results of the first test of the CUPID baseline module. In particular, we propose a new optimized detector structure and light sensors design to enhance the engineering and the light collection, respectively.\\
We characterized the heat detectors, achieving an energy resolution of (5.9 $\pm$ 0.2)\,keV FWHM at the $Q$-value of $^{100}$Mo (about 3034\,keV). 
We studied the light collection of the baseline CUPID design with respect to an alternative configuration which features gravity-assisted light detectors' mounting. In both cases we obtained an improvement in the light collection with respect to past measures and we validated the particle identification capability of the detector, which ensures an $\alpha$ particle rejection higher than 99.9\%, fully satisfying the requirements for CUPID.
\end{abstract}

\sloppy
\section{Introduction}
The two-neutrino double $\beta$ decay (2$\nu\beta\beta$) \cite{GoeppertMayer} is one of the rarest process in the universe, observed only in 11 nuclides, with typical half-lives in the range between 10$^{18}$ and 10$^{24}$\,yr \cite{BARABASH2020}. The precision measurements performed by several experiments allowed detailed studies of the 2$\nu\beta\beta$ spectral shape to search for distortions due to beyond Standard Model processes \cite{PhysRevD.100.092002,PhysRevLett.123.262501,PhysRevD.93.072001,nemo3-dbd,PhysRevLett.122.192501,PhysRevD.98.092007,Armengaud2020_2nu}. \\
An alternative mode to this process requires the emission of 2 electrons without neutrinos in the final state and is referred to as neutrinoless double $\beta$ decay (0$\nu\beta\beta$). This has been first hypothesised by W.H. Furry in 1939 \cite{Furry_1939} and then supported by several theoretical frameworks \cite{Deppisch:2012nb, Doi1985, Primakoff_1959, Mohapatra1986, VERGADOS19861}. 
The search for this decay plays a significant role in particle physics nowadays, as its discovery would establish the Dirac or Majorana nature of the neutrino, whose experimental evidence is still missing. In the former (Dirac) case, the neutrino behaves like all the other fermions. In the latter (Majorana) case, neutrino and antineutrino coincide, giving rise to new physics processes in which the total lepton number symmetry is violated, such as the 0$\nu\beta\beta$ decay \cite{Vissani:2016}. This would also represent an important hint for the explanation of the matter-antimatter asymmetry in the universe \cite{baryonasym, antimatter}. Several experiments have been searching for 0$\nu\beta\beta$ in different nuclides with sensitivities on the half-life from 10$^{24}$ to 10$^{26}$ yr \cite{Agostini1445,PhysRevLett.117.082503,PhysRevC.100.025501,PhysRevLett.123.161802,PhysRevD.92.072011,PhysRevLett.123.032501,PhysRevLett.124.122501,cupidmo2021,gerdacollaboration2020final} but still no evidence of this decay has been found.\\
The next generation experiment CUPID (CUORE Upgrade with Particle IDentification) \cite{CUPID_preCDR_2019} aims to explore the half-life region up to 10$^{27}$ years. CUPID will use scintillating cryogenic calorimeters, also called bolometers.
These are very low temperature detectors, operated at about 10 mK, whose main element is a crystal containing the isotope candidate for the 0$\nu\beta\beta$ emission. The crystal's heat capacity at cryogenic temperatures allows to convert an energy release into a measurable temperature increase. The temperature variation is then turned into an electric signal by means of a cryogenic sensor, called a thermistor.
This detection mechanism is the key to achieve an excellent energy resolution, about 0.2\,\% FWHM at a few MeV of energy deposit, which is one of the fundamental ingredients to increase the experimental sensitivity to the 0$\nu\beta\beta$ search. Moreover, these detectors feature a very high \DBD \  containment efficiency ($\sim$80\,\%) as the crystals work both as source and absorber of the decay products.
The CUPID experiment is based on years of development of such technology \cite{Fiorini:1983yj,pirroreview} culminated in the CUORE (Cryogenic Underground Observatory for Rare Events) experiment \cite{adams2021high,cuorecollaboration2021cuore}. By collecting more than 1 tonne-year of exposure in stable conditions, CUORE set a fundamental milestone for the next generation experiments searching for rare events with cryogenic calorimeters. Despite the many results achieved, CUORE is limited by the dominant background source of $\alpha$ particles produced by surface contaminations \cite{cuore-bkg-2017, Alduino_analysis_2016}. This dominant background source can be rejected by means of scintillating crystals with dual read-out of light and heat signals. Indeed, at a fixed energy deposit, the light yield of $\alpha$ particles is quenched with respect to $\beta/\gamma$ \cite{TRETYAK201040}. The particle identification represents the main innovation of the CUPID experiment, which will couple scintillating Li$_{2}${}$^{100}$MoO$_4$ crystals to light detectors to reject $\alpha$ events to a negligible level. Moreover, CUPID will search for the 0$\nu\beta\beta$ in the isotope $^{100}$Mo which presents an important feature; indeed, its $Q$-value, (3034.40 $\pm$ 0.17)\,keV \cite{RAHAMAN2008111}, lies above the last significant $\gamma$ line from natural radioactivity (at 2615\,keV) and this will further reduce the background level in the ROI, by mitigating the contribution due to $\gamma$s. \\
The combination of scintillating bolometers and high $Q$-value $\beta\beta$ emitters was exploited by LUCIFER \cite{Beeman:2012jd,Beeman:2012gg,Cardani:2013mja,Beeman:2013sba,Beeman:2013vda,Cardani_2013,Artusa:2016maw,Azzolini_2018} and LUMINEU \cite{BEEMAN2012318,Barabash:2014una,Armengaud:2015hda,Bekker:2014tfa,Armengaud_2017,Grigorieva2017,Poda2017} as well as by the AMoRE Collaboration \cite{AMORE2019}. The experience achieved in LUCIFER and LUMINEU resulted in two demonstrators which proved the CUPID working principles: CUPID-0 \cite{PhysRevD.100.092002,PhysRevLett.123.262501,PhysRevLett.123.032501,Azzolini_2019,Azzolini_excited_states,Azzolini_Zn_decay} and CUPID-Mo \cite{cupidmo2021,Poda2017,Armengaud2020,Schmidt_2020}. 
The former used cylindrical ZnSe scintillating crystals and Ge-disk bolometers as light detectors and took data from 2017 to 2020 at Laboratori Nazionali del Gran Sasso (LNGS).
CUPID-Mo operated at Laboratoire Souterrain de Modane (LSM) from early 2019 to mid 2020, proving the excellent radiopurity, energy resolution and $\alpha$ particles rejection achieved with cylindrical crystals of Li$_{2}${}$^{100}$MoO$_4$ (LMO), the compound chosen for CUPID. \\
The ongoing R\&D measurements at the LNGS \cite{hallc2020, pileup2021} and at the Canfranc laboratories \cite{cross2021} aim to optimize the detector features and design for the CUPID experiment. In this work we propose a new mechanical structure for the assembly of the CUPID baseline detector module and present its performance.

\section{Experimental Setup}
\begin{figure}[htbp]
\centering
\includegraphics[scale=0.48]{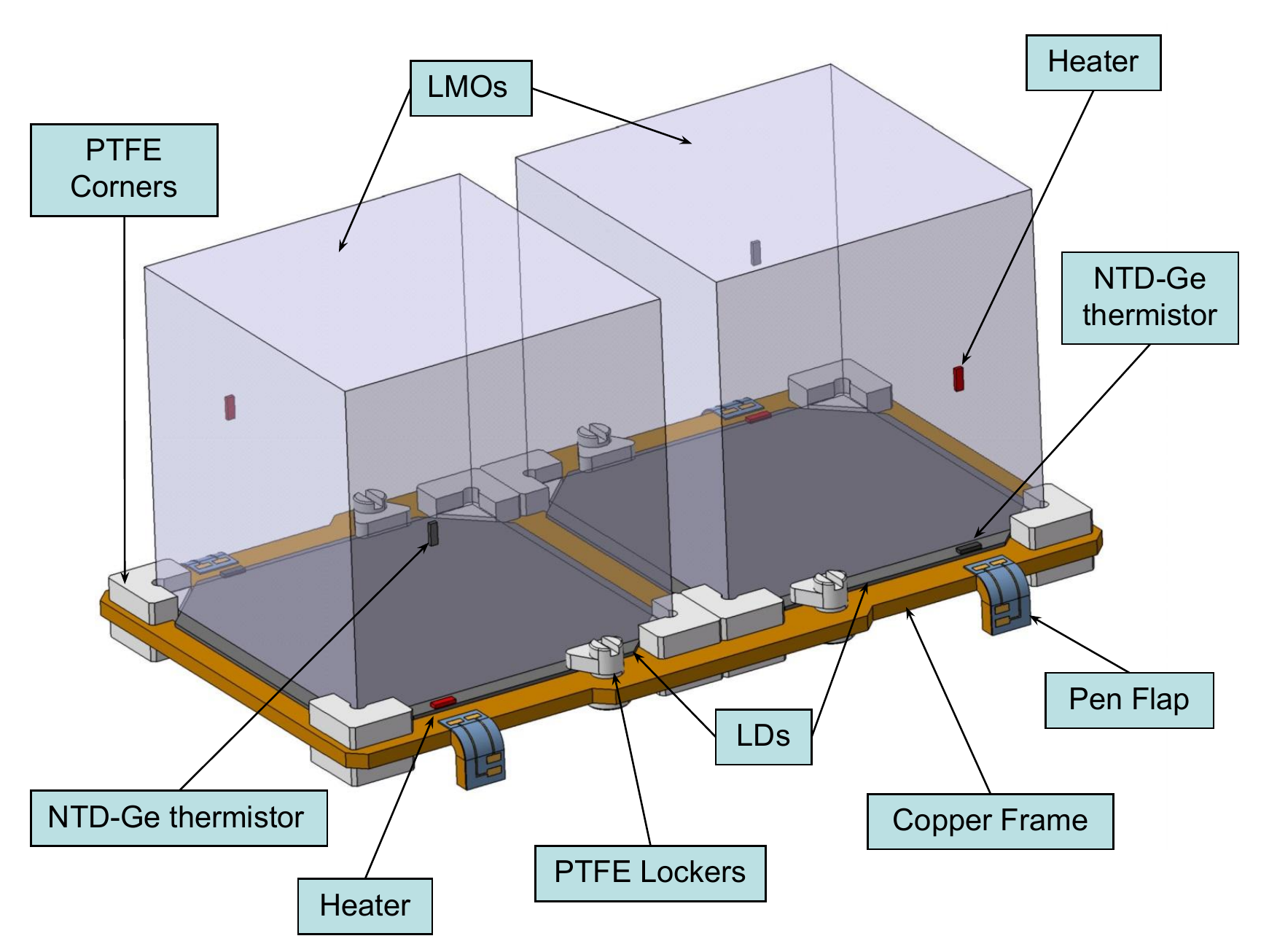}
\caption{Rendering of a single CUPID module which consists of 2 cubic LMO crystals and 2 LDs spaced 0.5\,mm from the bottom faces. The detectors are held by the copper frame and PTFE elements. A tower is built by simply stacking one module on top of the other. The detector components are labeled in the figure.}
\label{singlemodule}
\end{figure}

All the past/present bolometric detectors, from CUORE to CUPID-0 and CUPID-Mo, were assembled by mounting the crystals into copper frames, that were rigidly secured on top of each others using copper columns \cite{CUORE:2016aqq, CUPID:2018kff, Armengaud2020}.
Experiments that featured light detectors (LDs) in their setup, usually mounted these devices by ``squeezing" them into polytetrafluoroethylene (PTFE) clamps.
To simplify the CUPID assembly, we designed a new mechanical structure, in which two LDs are mounted into a 2\,mm-thick laser-cut copper frame (Fig. \ref{singlemodule}). 

\begin{figure}[htbp]
\centering
\includegraphics[scale=0.57]{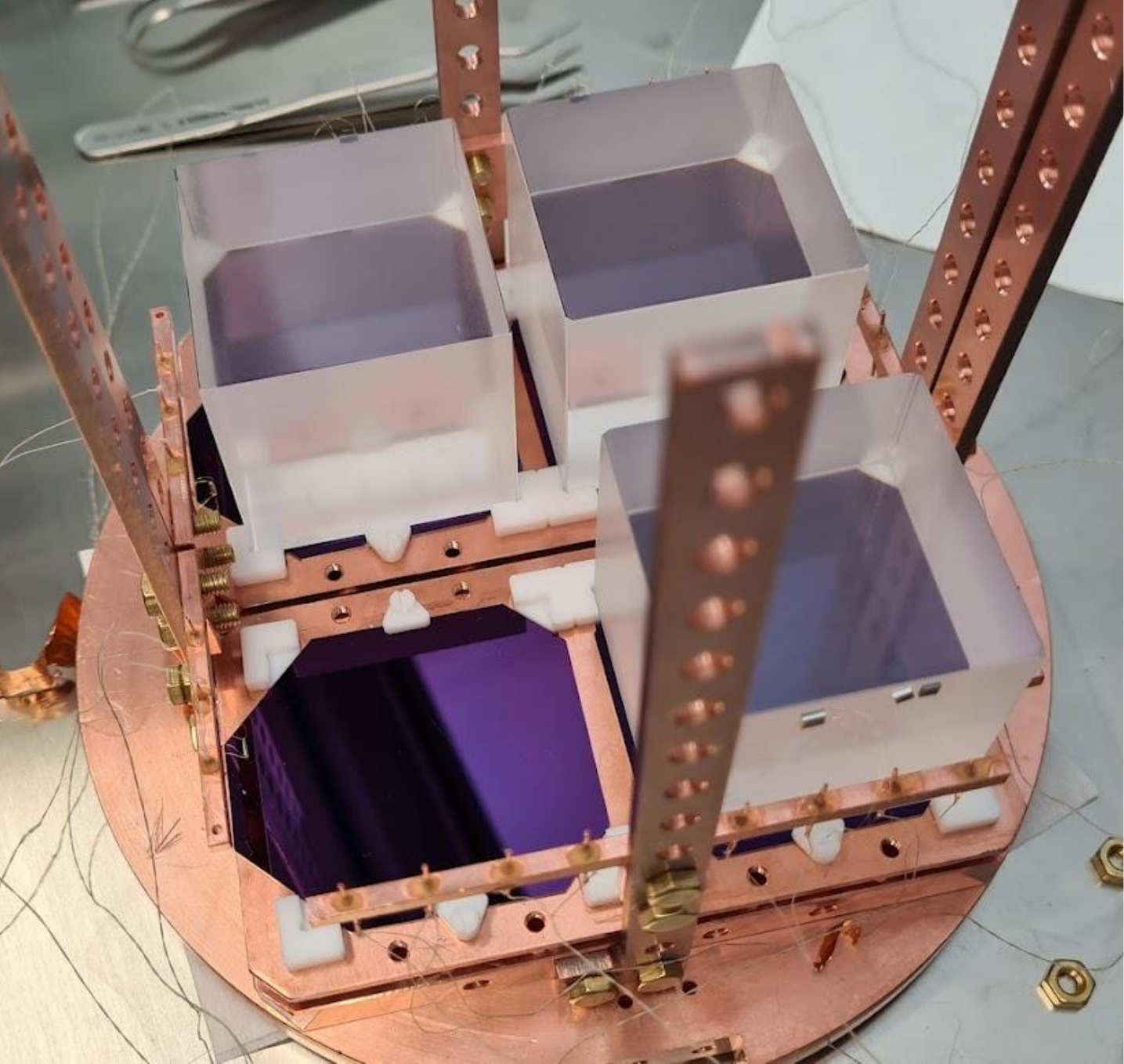}
\caption{Photo of the array during the decommissioning. Here are shown 2 baseline modules of CUPID, which form the first floors of the 2 mini-towers. In one of the 2 modules an LMO crystal is missing and a quasi-square LD is visible on the bottom.}
\label{photo}
\end{figure}

Instead of squeezing the LDs in the clamps, we keep them positioned on the edges of the copper frame using two PTFE ``lockers". Furthermore, LDs were re-designed to match the CUPID crystals faces; previous measurements used CUPID-Mo \cite{Armengaud2020} and CUPID-0 \cite{Beeman_2013} light detectors, consisting of 170\,$\mu$m thick, disk-shaped Ge LDs. We replaced the disk-shaped LDs with quasi-square ones and, to relax the constraints on the tolerances of PTFE elements, we also increased the thickness of the LDs from 170\,$\mu$m to 500\,$\mu$m. We verified that the effect of the volume increase on the heat capacity does not affect the time response.\\
The copper frame is equipped with PTFE ``corners" to ease the positioning of the LMO crystals as close as possible to the LDs (0.5\,mm, largely improving the 4\,mm spacing that could be achieved with the previous assembly procedure). Finally, the module design includes a pen flap, glued on the copper frame, to allow the bonding of Neutron Transmutation Doped germanium (NTD-Ge \cite{Hallerf}) thermistors and the heaters (P-doped Si \cite{Andreotti2012}). The former are meant for the read-out of both the LMO crystals and LDs, and produce a typical voltage signal of  10 -- 100 $\mu$V per 1\,MeV of deposited energy. The latter are used to periodically inject thermal pulses at a fixed energy for the thermal gain correction \cite{CUORE:2016aqq}. \\
A CUPID tower will consist of 14 modules stacked on top of each others simply by gravity. 
This structure shows several advantages: it minimizes the amount of inert material, relaxes the constraints on mechanical tolerances, simplifies the production of copper elements and their cleaning, and also the assembly is simplified as well. On the other hand, such novel design was never tested before, so the thermal properties and the propagation of noise across the floors of the tower needed to be characterized. Furthermore, the novel assembly of the LDs could potentially induce a larger noise, due to a less firm positioning of the detector module components with respect to previous assemblies.
In this work we made an exploratory study of the new mechanical structure, by mounting 2 mini-towers of only 2 (out of 14) floors each (Fig. \ref{photo}).

\begin{figure}[htbp]
\centering
\includegraphics[scale=0.48]{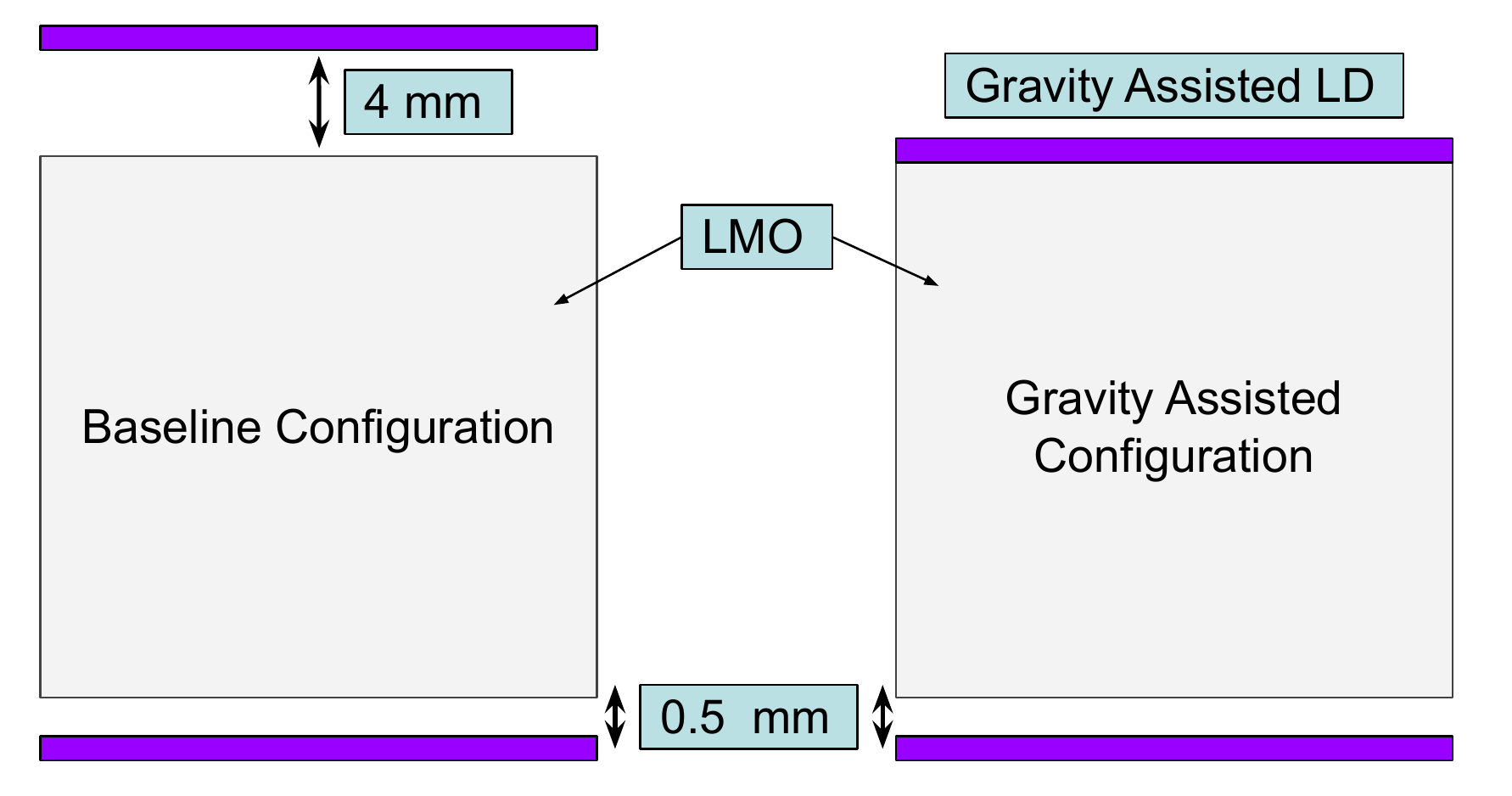}
\caption{Schematic view of the 2 LD configurations: the grey squares and the purple strips represent respectively the LMO crystals and the LDs. Left: ``baseline" configuration with LDs spaced 0.5 and 4\,mm from the bottom and top of the crystal respectively. Right: ``gravity-assisted" configuration with LDs spaced 0.5\,mm on bottom and leaned on top of the crystal.}
\label{scheme}
\end{figure}

Each floor hosted 2 natural LMO cubic crystals, with dimensions 45$\times$45$\times$45\,mm$^3$ and mass $\sim$280\,g each, for a total of 8 crystals with the same specifications foreseen for CUPID. High radiopurity copper and PTFE elements were selected for the mechanical structure.
Each crystal faces 2 LDs, on top and bottom. These are thin cryogenic germanium calorimeters. 
An antireflecting 60\,nm thick layer of SiO \cite{Mancuso2014} was deposited on both sides of the Ge faces to increase the light collection, as already done in CUPID-Mo \cite{Armengaud2020}. 
This is the most reliable technology to be operated at cryogenic temperatures 
to detect scintillation light from the crystals, which typically corresponds to an equivalent deposition of keV per MeV of energy deposit in the crystal \cite{Armengaud_2017,Poda2017,Armengaud2020,hallc2020,podascintillation}. 
\\
We tested 2 possible configurations of the LDs as outlined in Fig. \ref{scheme}: 
\begin{itemize}
    \item in the first floor, LDs were spaced 0.5 and 4\,mm from the bottom and top of the crystal respectively. From now on, we will refer to this configuration as the ``baseline" configuration for the CUPID experiment, as it allows a simple engineering of a large modular array.
    \item in the second floor, the bottom LDs were again spaced 0.5\,mm, but the top LDs were leaned on the LMO crystals. This ``gravity-assisted" positioning, originally proposed in Ref. \cite{Barucci_2019}, could allow to further increase the light collection.
\end{itemize}

One of the goals of this measurement was to establish if the larger light collection which could be offered by the ``gravity-assisted" configuration is worth the complication in the detector engineering and assembly.
We performed two experimental runs, one in the configuration described above, and the second one by surrounding the crystals with a Vikuiti{\texttrademark} reflecting foil (a potential back-up solution to further enhance the light collection).

The LDs were constantly exposed to an X-rays source ($^{55}$Fe, which produces peaks at 5.9 and 6.4\,keV) to energy-calibrate the scintillation light signals. \\
The prototype was operated in a wet cryostat located in the Hall C of the deep underground Laboratori Nazionali del Gran Sasso of INFN, Italy.

\section{Data Analysis}
\label{analysis}
The voltage signals from the detectors were amplified and filtered with a 120 dB/decade, six-pole anti-aliasing active Bessel filter~\cite{Arnaboldi_2018,Carniti2016,Arnaboldi_2008,Arnaboldi2015,Arnaboldi2010,Arnaboldi:2004jj,AProgFE}. We used a custom DAQ software package to save on disk the data stream acquired through a 18\,bit analog-to-digital board with a sampling frequency of 2\,kHz~\cite{DiDomizio:2018ldc}. Then, a derivative trigger \cite{Branca_2020} was applied to the data, to identify thermal pulses, and a random trigger was fired every 60\,s to sample the noise waveforms. The trigger parameters were tuned for each detector to optimize the noise level. We acquired heat and light pulses with a 5\,s and 0.5\,s long window respectively. \\
The triggered data were then processed offline via a dedicated analysis chain, which was adapted from a C++ based analysis framework developed for CUORE \cite{Alduino_analysis_2016}, CUPID-0~\cite{Azzolini:analysis:2018} and their predecessors~\cite{Andreotti_2011}.
The first step of the analysis was the application of a matched filter algorithm (optimum filter) ~\cite{Gatti:1986cw,Radeka:1966} to enhance the signal-to-noise ratio suppressing the most intense noise frequencies. This algorithm takes as input an average pulse and an average noise power spectrum, computed from recorded signal and noise waveforms, after a quality selection cut. The filter allows to improve the reconstruction of the basic pulse characteristics such as the amplitude, the baseline value (which is a proxy for the temperature), the baseline RMS and the pulse shape parameters. The filter was applied to both LD and LMO events. \\
Any time the trigger of an LMO crystal fired, the waveform of the corresponding LDs was acquired and flagged as ``side pulses". We exploited the fixed time delay between light and heat pulses due to the electronics in order to improve the estimate of the side pulses amplitude, which presents a poor signal-to-noise ratio at low energies. We estimated this fixed time delay for each LD from the average pulse corresponding to each channel and we evaluated the light pulses amplitude at the exact time delay with respect to the corresponding heat pulse. This allows to remove some non-linearities introduced by the optimum filter at low energies, while it does not affect significantly the light signals amplitude in the region of interest~\cite{Azzolini:analysis:2018}.\\
The light signals amplitudes were energy-calibrated by using a linear function with zero intercept. The calibration coefficient was derived by fitting the peaks at 5.9 and 6.4\,keV  of the $^{55}$Fe source. \\
Concerning the heat channel, a further improvement of the pulses amplitude estimation was possible through a thermal gain correction.
Unlike the LDs, which exhibit typical resolutions of $\sim$1\% \cite{cross2021}, the heat channels are expected to reach a resolution of a few $\sim$0.1\%. Such energy resolution could be spoiled by thermal instabilities of the cryostat.
During the data taking we placed a $^{232}$Th source outside the cryostat to derive the amplitude vs. temperature dependence in the highest energy peak from $^{208}$Tl at 2615 keV, and correct the pulse amplitude accordingly. 
Finally, to convert the corrected amplitudes into energy, we identified and fitted the most intense mono-energetic peaks from the $^{232}$Th decay chain. Then, we calibrated the heat signals by using a second order polynomial function crossing the origin, which showed residuals lower than $\sim$1\,keV in absolute value.

\section{General Performances}
The temperature at which we operated the detectors of the array is $\sim$15\,mK.
The results presented in this paper are focused on the LDs which showed the best performances. We operated the LDs with working resistances in the range of 4--7 M${\rm \Omega}$, to obtain a response spanning from 0.9 to 5.2\,$\mu$V/keV (average $\sim$2.7\,$\mu$V/keV), depending on the bias current.
\\
We evaluated the baseline RMS from a Gaussian fit to the energy spectrum of noise events and found a result between 35 and 70\,eV with an average of (57 $\pm$ 6)\,eV, well below the threshold required for CUPID (100 eV) \cite{CUPID_preCDR_2019}.\\
We did not find any correlation between the LD performance and its position in the tower. This confirms the homogeneity of the cooling along the tower and the uniformity of the results among the different LDs assembly methods. The validation of the new detector structure used in this measurement is the first important result towards the successful construction of a 14-floors tower prototype for the CUPID experiment.\\

We operated the LMO crystals with working resistances of 2--13\,M${\rm \Omega}$ and we measured the response to be in the range 31--72\, $\mu$V/MeV (average $\sim$50\,$\mu$V/MeV).
We evaluated the baseline resolution as done for the LDs, with a result between 0.52 and 0.95\,keV RMS depending on the LMO detector, with an overall average of (0.69 $\pm$ 0.06)\,keV RMS. The obtained performance is in agreement with previous cubic LMO detectors tested in the same facility \cite{hallc2020} and consistent with cylindrical and cubic LMO detectors operated in similar conditions \cite{Armengaud_2017,Armengaud2020,cross2021}. 

\begin{figure}[htbp]
\centering
\includegraphics[scale=0.43]{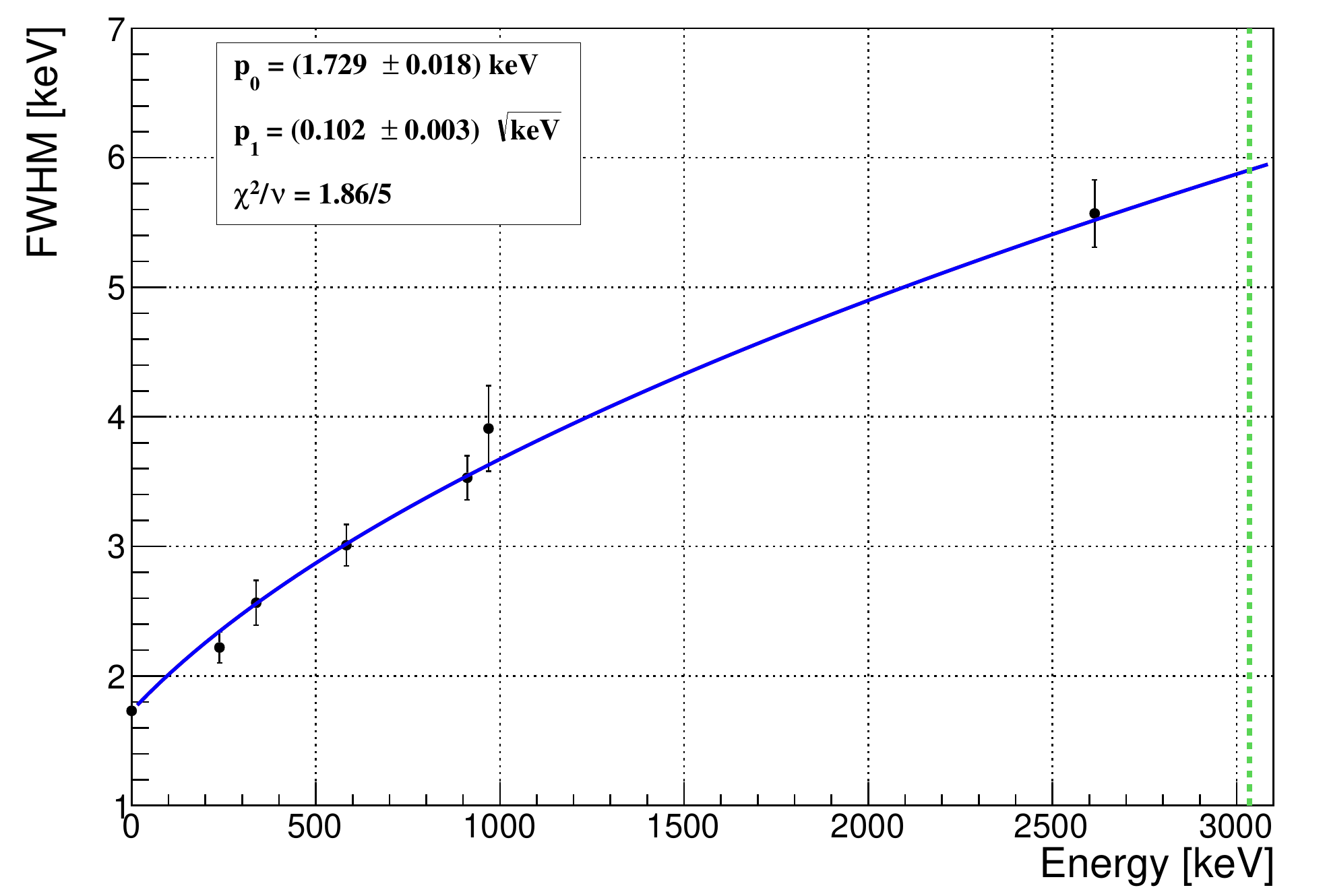}
\caption{FWHM of the most intense $\gamma$ peaks as a function of the energy in the sum spectrum of the 8 LMO crystals. We find that the function which best describes the data is FWHM = $\sqrt{p_0^2 + p_1^2 \times E}$, here shown with a blue line. The green dotted line represents the $^{100}$Mo $Q$-value.}
\label{reso}
\end{figure}

We evaluated the energy resolution at different energies by fitting the most intense $\gamma$ peaks in the sum spectrum of the 8 LMO crystals (Fig.~\ref{reso}).

We find a resolution of (5.6 $\pm$ 0.3)\,keV FWHM at the 2615 keV $^{208}$Tl peak. 
To extrapolate the energy resolution at the $Q_{\beta\beta}$ of $^{100}$Mo (3034 keV), 
we performed different tests to evaluate the best function to describe the energy dependence of the energy resolution. We find that a square root function with a linear dependence on energy, namely $\sqrt{p_0^2 + p_1^2 \times E}$, is the best fit to the data (Fig.~\ref{reso}). We also included in the fit the baseline resolution estimated on noise events from the overall spectrum, to better constrain the $p_0$ parameter. 
The extrapolated FWHM at the $Q_{\beta\beta}$ of $^{100}$Mo is (5.9 $\pm$ 0.2) keV, which corresponds to a percentage resolution of 0.19\%. 
The improved signal-to-noise ratio, due to an extensive characterization of the noise sources of this facility, allowed us to improve the energy resolution compared to the previous tests~\cite{hallc2020}, approaching the final CUPID goal of 5\,keV FWHM.\\

\section{Light Collection Results}
One of the main purposes of the run was the evaluation of the light collection of quasi-square light detectors and a comparison of the results obtained with different spacing between LDs and crystals. To estimate the total light collection, we added the corresponding light amplitude of top and bottom LDs and divided by the energy estimated from the heat channel. The total light yield (LY) is then calculated as the mean of the resulting distribution. 

\begin{figure}[htbp]
\centering
\includegraphics[scale=0.43]{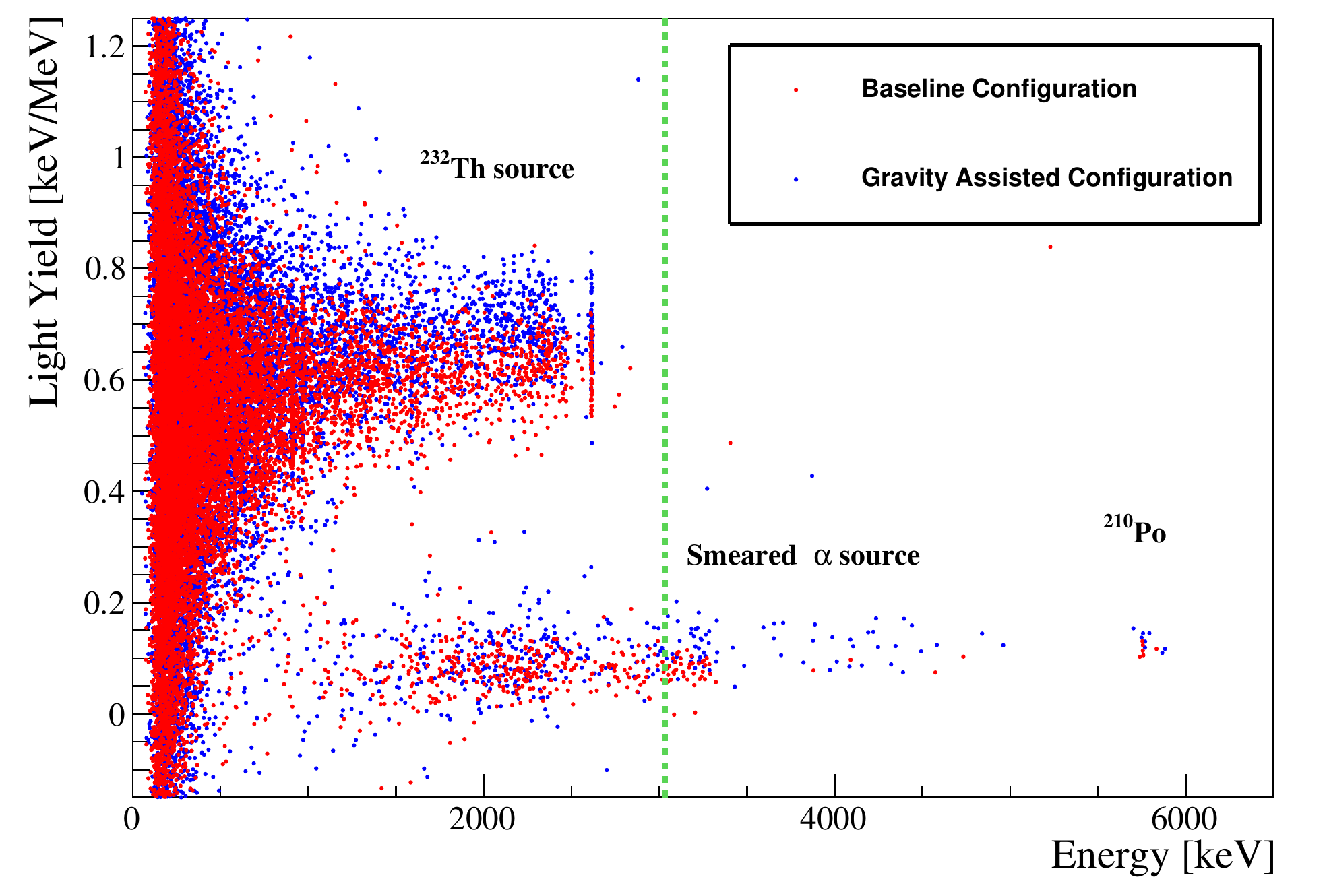}
\caption{Light Yield as a function of the energy deposited in the LMO crystal. Red: baseline configuration; blue: gravity-assisted configuration. Green vertical line: $Q$-value of $^{100}$Mo.}
\label{ly}
\end{figure}

We report the total LY as a function of energy for two crystals representative of the two configurations (baseline and gravity-assisted) in Fig.~\ref{ly}. The results obtained with the other crystals of the respective configuration are very similar, with differences smaller than 20\,\%. 

In both configurations we can clearly identify the $\beta/\gamma$ events, which populate the plot up to the $^{208}$Tl line at 2615\,keV, and the $\alpha$ events, which present a quenched light yield and extend up to higher energies. In particular, we identify a cluster of events due to an internal crystal contamination in $^{210}$Po~\cite{Armengaud:2015hda}, which produces a peak with nominal energy $\sim$5.4\,MeV. Since the detector was energy-calibrated using gamma’s, the $\alpha$ peak is observed at slightly higher electron-equivalent energy (+7\,\%, in agreement with previous studies with lithium molybdate bolometers \cite{Cardani_2013,Bekker:2014tfa,Armengaud_2017,Armengaud2020,clymene:2018}). The $\alpha$ events at lower energies are produced by a $^{234}$U/$^{238}$U source covered with a thin Mylar foil to smear the energy of $\alpha$ particles, to study the light collection in the ROI for the 0$\nu\beta\beta$ search. The LY distribution shows a spread at very low energies due to the superposition of the noise with the light pulses. For this reason, to avoid the impact of noise on the LY estimation, we selected scintillation events with energy deposit in the crystal above 1.2\,MeV.

The average total LY$_{\beta/\gamma}$ is found to be (0.62 $\pm$ 0.04)\,keV/MeV and (0.70 $\pm$ 0.05)\,keV/MeV in the ``baseline" configuration and in the ``gravity-assisted" configuration, respectively. 

In particular, for the ``baseline" configuration, the LYs of a single LD resulted to be on average (0.28 $\pm$ 0.02)\,keV/MeV for the LD spaced 4\,mm and (0.33 $\pm$ 0.03)\,keV/MeV for the LD spaced 0.5\,mm. In the ``gravity-assisted" configuration we found the LY of a single LD to be (0.36 $\pm$ 0.03)\,keV/MeV for both the LDs. More details on the LYs of a single LD are reported in Table \ref{Table:LY-features-2}.

\begin{table}[htbp]
\centering
\caption{Light yield for LD top (t), LD bottom (b) and the sum of the two light detectors in the case of bare crystals.
LMO-1 to LMO-4 are in the baseline configuration (bottom LD spaced 0.5\,mm and top LD spaced 4\,mm). LMO-5 to LMO-8 are in the ``gravity assisted" configuration (bottom LD spaced 0.5\,mm and top LD leaned on crystal). The missing values correspond to LDs we discarded for the analysis (LMO-1 top LD corresponds to LMO-5 bottom LD). The associated uncertainty is about 10\% on each value.}
\begin{tabular}{lccc}
\hline
                       &   LY$_{\beta/\gamma}$ (t)                 &  LY$_{\beta/\gamma}$ (b)  &   LY$_{\beta/\gamma}$ (sum)        \\
                      &[keV/MeV]        &[keV/MeV]           &[keV/MeV]                           \\
\hline
\hline
LMO-1           & -                            & 0.35             	 & -          \\
\hline
LMO-2           & 0.29                            & 0.33             	 & 0.62          \\
\hline
LMO-3        & 0.26                             & -             	    & -       \\
\hline
LMO-4          & 0.30                               & 0.32                           & 0.63        \\
\hline
\hline
LMO-5         & 0.38                             & -            	  & -          \\
\hline
LMO-6         & 0.35                               & 0.35               	  &0.69       \\
\hline
LMO-7         & 0.34                               & 0.35                           & 0.69        \\
\hline
LMO-8         & 0.36                               & 0.37                           & 0.74        \\
\hline
\end{tabular}
\label{Table:LY-features-2}
\end{table}

The total LY for $\alpha$ particles resulted to be (0.08 $\pm$ 0.03)\,keV/MeV for the ``baseline" configuration and (0.11 $\pm$ 0.03)\,keV/MeV for the ``gravity-assisted" one. 

We repeated the same study on the prototype in which the LMO crystals were surrounded by reflecting foils, obtaining in both the configurations an increase of the LY by a factor 2, as already found in Ref. \cite{hallc2020}. \\

From these results we conclude that the increased complexity in engineering and mounting the LDs in the ``gravity-assisted" configuration is not motivated by a substantial gain in the light collection performance. For this reason, we decided to discard the ``gravity-assisted" configuration in view of CUPID and from now on we will focus on the ``baseline" configuration only.

To quantify the particle identification capabilities of the baseline configuration,  we define the Discrimination Power (DP) \cite{Arnaboldi2011797} as:
\begin{equation}
		DP \equiv \frac{\left| LY_{\beta/\gamma} - LY_{\alpha} \right|}{\sqrt{\sigma_{\beta/\gamma}^2 + \sigma_{\alpha}^2}}.
		\label{eq:discr_power}
\end{equation}

We find that the DP for the sum of LDs ranges between 7.3 and 8.2, thus largely exceeding the requirements of CUPID. Indeed, the minimum DP needed to reject the 99.9 \% of $\alpha$ particles is 3.1. In the unlikely event of a loss of a light detector, the DP would diminish. Assuming that a single LD is working, we obtain a DP between 3.9 and 6.2, thus closer but still higher than the required threshold.
It is worth noticing, in this context, that in the assembly of the CUORE detector only 4 out of 988 contacts were lost and that in the CUPID-0 detector none of the $\sim$30 LDs exhibited any malfunction. 

Finally, we compared our results with the R\&D test made on disk-shaped LDs and cubic LMO crystals in the same facility~\cite{hallc2020}. To allow a coherent comparison, we considered the results obtained with the ``baseline" configuration, in which the spacing between the LDs and the LMOs was similar to the assembly used in the previous test. The LY calculated from the light collected by both top and bottom LDs shows an average improvement of about 26\%. This factor is consistent with the improvement expected from increasing the geometrical size of the detector ($\sim$ 27\%). 

\section{Summary and Conclusions}
In view of the CUPID experiment, we modified the detector design to optimize its engineering and improve the light collection efficiency.\\
We validated the assembly of the new detector structure, which did not show any temperature gradient throughout the setup. This is a first fundamental step towards the construction of the 14-floors prototype tower of CUPID, which is planned for the first half of 2022. \\
We achieved a baseline resolution of (0.69 $\pm$ 0.06)\,keV RMS for the LMO crystals and we estimated an energy resolution of (5.9 $\pm$ 0.2)\,keV FWHM at the $Q$-value of $^{100}$Mo (3034\,keV).\\ 

To optimize the light collection we also redesigned the LDs to fully cover the faces of the CUPID crystals. For the first time, we characterized the performances of quasi-square LDs coupled to LMO cubic crystals. We estimated and demonstrated an improvement of the light collection of about 26\% with respect to disk-shaped LDs, which have been tested in the same facility. Moreover, the noise level achieved with the new LDs falls in the range 35--70\,eV which is well below the threshold required for CUPID (100\,eV). This pushes at limits the particle identification capabilities for CUPID, which guarantee an $\alpha$ particles rejection higher than 99.9\%.\\
Finally, we tested 2 possible configurations for the LD mounting. In the ``baseline" CUPID configuration, LDs are spaced 0.5\,mm and 4\,mm from the bottom and top faces of the crystal respectively. We also tested a ``gravity-assisted" configuration in which the top LD is leaned on the crystal. We eventually discarded the latter option as a viable solution for CUPID, since the gain in light collection is not worth the increased technical complexity of the assembly.

\section*{Acknowledgments}
The CUPID Collaboration thanks the directors and staff of the
Laboratori Nazionali del Gran Sasso and the technical staff of our
laboratories. This work was supported by the Istituto Nazionale di
Fisica Nucleare (INFN); by the European Research Council (ERC) under
the European Union Horizon 2020 program (H2020/2014-2020) with the ERC
Advanced Grant no. 742345 (ERC-2016-ADG, project CROSS) and the Marie
Sklodowska-Curie Grant Agreement No. 754496; by the Italian Ministry
of University and Research (MIUR) through the grant Progetti di
ricerca di Rilevante Interesse Nazionale (PRIN 2017, grant
no. 2017FJZMCJ); by the US National Science Foundation under Grant
Nos. NSF-PHY-1401832, NSF-PHY-1614611, and NSF-PHY-1913374. This material is also based
upon work supported by the US Department of Energy (DOE) Office of
Science under Contract Nos. DE-AC02-05CH11231 and DE-AC02-06CH11357;
and by the DOE Office of Science, Office of Nuclear Physics under
Contract Nos. DE-FG02-08ER41551, DE-SC0011091, DE-SC0012654,
DE-SC0019316, DE-SC0019368, and DE-SC0020423. This work was also
supported by the Russian Science Foundation under grant
No. 18-12-00003 and the National Research Foundation of Ukraine under
Grant No. 2020.02/0011. This research used resources of the National
Energy Research Scientific Computing Center (NERSC). This work makes
use of both the DIANA data analysis and APOLLO data acquisition
software packages, which were developed by the CUORICINO, CUORE,
LUCIFER and CUPID-0 Collaborations.

\bibliographystyle{spphys}       
\bibliography{hallC.bib}   

\begin{thebibliography}{10}
\providecommand{\url}[1]{{#1}}
\providecommand{\urlprefix}{URL }
\expandafter\ifx\csname urlstyle\endcsname\relax
  \providecommand{\doi}[1]{DOI \discretionary{}{}{}#1}\else
  \providecommand{\doi}{DOI \discretionary{}{}{}\begingroup
  \urlstyle{rm}\Url}\fi

\bibitem{GoeppertMayer}
M.~Goeppert-Mayer, Phys. Rev. \textbf{48}, 512 (1935).
\newblock \doi{10.1103/PhysRev.48.512}

\bibitem{BARABASH2020}
A.S. Barabash, Universe \textbf{2020}(6), 159 (2020).
\newblock \doi{10.3390/universe6100159}

\bibitem{PhysRevD.100.092002}
O.~Azzolini, et~al., Phys. Rev. D \textbf{100}, 092002 (2019).
\newblock \doi{10.1103/PhysRevD.100.092002}

\bibitem{PhysRevLett.123.262501}
O.~Azzolini, et~al., Phys. Rev. Lett. \textbf{123}, 262501 (2019).
\newblock \doi{10.1103/PhysRevLett.123.262501}

\bibitem{PhysRevD.93.072001}
J.B. Albert, et~al., Phys. Rev. D \textbf{93}, 072001 (2016).
\newblock \doi{10.1103/PhysRevD.93.072001}

\bibitem{nemo3-dbd}
R.~Arnold, et~al., Eur. Phys. J. C \textbf{79}(5), 440 (2019).
\newblock \doi{10.1140/epjc/s10052-019-6948-4}

\bibitem{PhysRevLett.122.192501}
A.~Gando, et~al., Phys. Rev. Lett. \textbf{122}, 192501 (2019).
\newblock \doi{10.1103/PhysRevLett.122.192501}

\bibitem{PhysRevD.98.092007}
A.S. Barabash, et~al., Phys. Rev. D \textbf{98}, 092007 (2018).
\newblock \doi{10.1103/PhysRevD.98.092007}

\bibitem{Armengaud2020_2nu}
E.~Armengaud, et~al., Eur. Phys. J. C \textbf{80}(7), 674 (2020).
\newblock \doi{10.1140/epjc/s10052-020-8203-4}

\bibitem{Furry_1939}
W.~Furry, Phys. Rev. \textbf{56}, 1184 (1939).
\newblock \doi{10.1103/PhysRev.56.1184}

\bibitem{Deppisch:2012nb}
F.F. Deppisch, M.~Hirsch, H.~P$\ddot{a}$s, J. Phys. G \textbf{39}, 124007
  (2012).
\newblock \doi{10.1088/0954-3899/39/12/124007}

\bibitem{Doi1985}
M.~Doi, T.~Kotani, E.~Takasugi, Progress of Theoretical Physics Supplement
  \textbf{83}, 1 (1985).
\newblock \doi{10.1143/PTPS.83.1}

\bibitem{Primakoff_1959}
H.~Primakoff, S.P. Rosen, Reports on Progress in Physics \textbf{22}(1), 121
  (1959).
\newblock \doi{10.1088/0034-4885/22/1/305}

\bibitem{Mohapatra1986}
R.N. Mohapatra, Phys. Rev. D \textbf{34}, 3457 (1986).
\newblock \doi{10.1103/PhysRevD.34.3457}

\bibitem{VERGADOS19861}
J.~Vergados, Physics Reports \textbf{133}(1), 1 (1986).
\newblock \doi{10.1016/0370-1573(86)90088-8}

\bibitem{Vissani:2016}
S.~Dell'Oro, et~al., Advances in High Energy Physics \textbf{2016}, 2162659
  (2016).
\newblock \doi{10.1155/2016/2162659}

\bibitem{baryonasym}
F.F. Deppisch, L.~Graf, J.~Harz, W.C. Huang, Phys. Rev. D \textbf{98}, 055029
  (2018).
\newblock \doi{10.1103/PhysRevD.98.055029}

\bibitem{antimatter}
M.~Fukugita, T.~Yanagida, Physics Letters B \textbf{174}(1), 45 (1986).
\newblock \doi{10.1016/0370-2693(86)91126-3}

\bibitem{Agostini1445}
M.~Agostini, et~al., Science \textbf{365}(6460), 1445 (2019).
\newblock \doi{10.1126/science.aav8613}

\bibitem{PhysRevLett.117.082503}
A.~Gando, et~al., Phys. Rev. Lett. \textbf{117}, 082503 (2016).
\newblock \doi{10.1103/PhysRevLett.117.082503}

\bibitem{PhysRevC.100.025501}
S.I. Alvis, et~al., Phys. Rev. C \textbf{100}, 025501 (2019).
\newblock \doi{10.1103/PhysRevC.100.025501}

\bibitem{PhysRevLett.123.161802}
G.~Anton, et~al., Phys. Rev. Lett. \textbf{123}, 161802 (2019).
\newblock \doi{10.1103/PhysRevLett.123.161802}

\bibitem{PhysRevD.92.072011}
R.~Arnold, et~al., Phys. Rev. D \textbf{92}, 072011 (2015).
\newblock \doi{10.1103/PhysRevD.92.072011}

\bibitem{PhysRevLett.123.032501}
O.~Azzolini, et~al., Phys. Rev. Lett. \textbf{123}, 032501 (2019).
\newblock \doi{10.1103/PhysRevLett.123.032501}

\bibitem{PhysRevLett.124.122501}
D.Q. Adams, et~al., Phys. Rev. Lett. \textbf{124}, 122501 (2020).
\newblock \doi{10.1103/PhysRevLett.124.122501}

\bibitem{cupidmo2021}
E.~Armengaud, et~al., Phys. Rev. Lett. \textbf{126}, 181802 (2021).
\newblock \doi{10.1103/physrevlett.126.181802}

\bibitem{gerdacollaboration2020final}
M.~Agostini, et~al., Phys. Rev. Lett. \textbf{125}, 252502 (2020).
\newblock \doi{10.1103/PhysRevLett.125.252502}

\bibitem{CUPID_preCDR_2019}
W.~Armstrong, et~al.,   (2019).
\newblock \urlprefix\url{https://arxiv.org/abs/1907.09376}

\bibitem{Fiorini:1983yj}
E.~Fiorini, T.O. Niinikoski, Nucl. Instrum. Meth. A \textbf{224}, 83 (1984)

\bibitem{pirroreview}
S.~Pirro, P.~Mauskopf, Annu. Rev. Nucl. Part. Sci. \textbf{67}, 161 (2017)

\bibitem{adams2021high}
D.Q. Adams, et~al.,   (2021).
\newblock \urlprefix\url{https://arxiv.org/abs/2104.06906}

\bibitem{cuorecollaboration2021cuore}
D.Q. Adams, et~al., Prog. Part. Nucl. Phys. \textbf{122}, 103902 (2022).
\newblock \doi{10.1016/j.ppnp.2021.103902}

\bibitem{cuore-bkg-2017}
C.~Alduino, et~al., Eur. Phys. J. C \textbf{77}(8), 543 (2017).
\newblock \doi{10.1140/epjc/s10052-017-5080-6}

\bibitem{Alduino_analysis_2016}
C.~Alduino, et~al., Phys. Rev. C \textbf{93}, 045503 (2016).
\newblock \doi{10.1103/physrevc.93.045503}

\bibitem{TRETYAK201040}
V.I. Tretyak, Astropart. Phys. \textbf{33}(1), 40 (2010).
\newblock \doi{10.1016/j.astropartphys.2009.11.002}

\bibitem{RAHAMAN2008111}
S.~Rahaman, et~al., Physics Letters B \textbf{662}(2), 111  (2008).
\newblock \doi{10.1016/j.physletb.2008.02.047}

\bibitem{Beeman:2012jd}
J.W. Beeman, et~al., Astropart. Phys. \textbf{35}, 813 (2012).
\newblock \doi{10.1016/j.astropartphys.2012.02.013}

\bibitem{Beeman:2012gg}
J.W. Beeman, et~al., Eur. Phys. J. C \textbf{72}, 2142 (2012).
\newblock \doi{10.1140/epjc/s10052-012-2142-7}

\bibitem{Cardani:2013mja}
L.~Cardani, et~al., J. Phys. G \textbf{41}, 075204 (2014).
\newblock \doi{10.1088/0954-3899/41/7/075204}

\bibitem{Beeman:2013sba}
J.W. Beeman, et~al., Adv. High Energy Phys. \textbf{2013}, 237973 (2013).
\newblock \doi{10.1155/2013/237973}

\bibitem{Beeman:2013vda}
J.W. Beeman, et~al., JINST \textbf{8}, P05021 (2013).
\newblock \doi{10.1088/1748-0221/8/05/P05021}

\bibitem{Cardani_2013}
L.~Cardani, et~al., JINST \textbf{8}, P10002 (2013).
\newblock \doi{10.1088/1748-0221/8/10/p10002}

\bibitem{Artusa:2016maw}
D.R. Artusa, et~al., Eur. Phys. J. C \textbf{76}(7), 364 (2016).
\newblock \doi{10.1140/epjc/s10052-016-4223-5}

\bibitem{Azzolini_2018}
O.~Azzolini, et~al., Eur. Phys. J. C \textbf{78}(5), 428 (2018).
\newblock \doi{10.1140/epjc/s10052-018-5896-8}

\bibitem{BEEMAN2012318}
J.~Beeman, et~al., Physics Letters B \textbf{710}(2), 318  (2012).
\newblock \doi{10.1016/j.physletb.2012.03.009}

\bibitem{Barabash:2014una}
A.~Barabash, et~al., Eur. Phys. J. C \textbf{74}(10), 3133 (2014).
\newblock \doi{10.1140/epjc/s10052-014-3133-7}

\bibitem{Armengaud:2015hda}
E.~Armengaud, et~al., JINST \textbf{10}, P05007 (2015).
\newblock \doi{10.1088/1748-0221/10/05/P05007}

\bibitem{Bekker:2014tfa}
T.B. Bekker, et~al., Astropart. Phys. \textbf{72}, 38 (2016).
\newblock \doi{10.1016/j.astropartphys.2015.06.002}

\bibitem{Armengaud_2017}
E.~Armengaud, et~al., Eur. Phys. J. C \textbf{77}(11), 785 (2017).
\newblock \doi{10.1140/epjc/s10052-017-5343-2}

\bibitem{Grigorieva2017}
V.~Grigorieva, et~al., J. Materials Science Engineering B \textbf{7}, 63
  (2017).
\newblock \doi{10.17265/2161-6221/2017.3-4.002}

\bibitem{Poda2017}
D.V. Poda, AIP Conference Proceedings \textbf{1894}(1), 020017 (2017).
\newblock \doi{10.1063/1.5007642}

\bibitem{AMORE2019}
V.~Alenkov, et~al., Eur. Phys. J. C \textbf{79}(9), 791 (2019).
\newblock \doi{10.1140/epjc/s10052-019-7279-1}

\bibitem{Azzolini_2019}
O.~Azzolini, et~al., Eur. Phys. J. C \textbf{79}(7), 583 (2019).
\newblock \doi{10.1140/epjc/s10052-019-7078-8}

\bibitem{Azzolini_excited_states}
O.~Azzolini, et~al., Eur. Phys. J. C \textbf{78}(11), 888 (2018).
\newblock \doi{10.1140/epjc/s10052-018-6340-9}

\bibitem{Azzolini_Zn_decay}
O.~Azzolini, et~al., Eur. Phys. J. C \textbf{80}(8), 702 (2020).
\newblock \doi{10.1140/epjc/s10052-020-8280-4}

\bibitem{Armengaud2020}
E.~Armengaud, et~al., Eur. Phys. J. C \textbf{80}(1), 44 (2020).
\newblock \doi{10.1140/epjc/s10052-019-7578-6}

\bibitem{Schmidt_2020}
B.~Schmidt, et~al., Journal of Physics: Conference Series \textbf{1468}, 012129
  (2020).
\newblock \doi{10.1088/1742-6596/1468/1/012129}

\bibitem{hallc2020}
A.~Armatol, et~al., Eur. Phys. J. C \textbf{81}, 104 (2021).
\newblock \doi{10.1140/epjc/s10052-020-08809-8}

\bibitem{pileup2021}
A.~Armatol, et~al., Phys. Rev. C \textbf{104}, 015501 (2021).
\newblock \doi{10.1103/physrevc.104.015501}

\bibitem{cross2021}
A.~Armatol, et~al., JINST \textbf{16}, P02037 (2021).
\newblock \doi{10.1088/1748-0221/16/02/p02037}

\bibitem{CUORE:2016aqq}
C.~Alduino, et~al., JINST \textbf{11}, P07009 (2016).
\newblock \doi{10.1088/1748-0221/11/07/P07009}

\bibitem{CUPID:2018kff}
O.~Azzolini, et~al., Eur. Phys. J. C \textbf{78}(5), 428 (2018).
\newblock \doi{10.1140/epjc/s10052-018-5896-8}

\bibitem{Beeman_2013}
J.W. Beeman, et~al., JINST \textbf{8}, P07021 (2013).
\newblock \doi{10.1088/1748-0221/8/07/p07021}

\bibitem{Hallerf}
E.E. Haller, N.P. Palaio, W.L. Hansen, E.~Kreysa, in: R.D. Larrabee~(ed.),
  \emph{Neutron Transmutation Doping of Semiconductor Materials} (Plenum Press,
  1984), p. 21.

\bibitem{Andreotti2012}
E.~{Andreotti}, et~al., Nucl. Instrum. Meth. A \textbf{664}(1), 161 (2012).
\newblock \doi{10.1016/j.nima.2011.10.065}

\bibitem{Mancuso2014}
M.~Mancuso, et~al., EPJ Web of Conferences \textbf{65}, 04003 (2014).
\newblock \doi{10.1051/epjconf/20136504003}

\bibitem{podascintillation}
D.~Poda, Physics \textbf{3}(3), 473 (2021).
\newblock \doi{10.3390/physics3030032}

\bibitem{Barucci_2019}
M.~Barucci, et~al., Nucl. Instrum. Meth. A \textbf{935}, 150 (2019).
\newblock \doi{10.1016/j.nima.2019.05.019}

\bibitem{Arnaboldi_2018}
C.~Arnaboldi, et~al., JINST \textbf{13}(02), P02026 (2018).
\newblock \doi{10.1088/1748-0221/13/02/p02026}

\bibitem{Carniti2016}
P.~Carniti, et~al., Rev. Sci. Instr. \textbf{87}, 054706 (2016).
\newblock \doi{10.1063/1.4948390}

\bibitem{Arnaboldi_2008}
C.~Arnaboldi, G.~Pessina, Journal of Low Temperature Physics \textbf{151}(3),
  964 (2008).
\newblock \doi{10.1007/s10909-008-9785-7}

\bibitem{Arnaboldi2015}
C.~Arnaboldi, et~al., Rev. Sci. Instr. \textbf{86}(12), 124703 (2015).
\newblock \doi{10.1063/1.4936269}

\bibitem{Arnaboldi2010}
C.~{Arnaboldi}, et~al., Nucl. Instrum. and Meth. A \textbf{617}(1-3), 327
  (2010).
\newblock \doi{10.1016/j.nima.2009.09.023}

\bibitem{Arnaboldi:2004jj}
C.~Arnaboldi, et~al., Nucl. Instrum. Meth. A \textbf{520}, 578 (2004).
\newblock \doi{10.1016/j.nima.2003.11.319}

\bibitem{AProgFE}
C.~Arnaboldi, et~al., IEEE Trans. Nucl. Sci. \textbf{49}(5), 2440 (2002).
\newblock \doi{0.1109/TNS.2002.803886}

\bibitem{DiDomizio:2018ldc}
S.D. Domizio, et~al., JINST \textbf{13}, P12003 (2018).
\newblock \doi{10.1088/1748-0221/13/12/p12003}

\bibitem{Branca_2020}
A.~Branca, Journal of Physics: Conference Series \textbf{1468}(1), 012118
  (2020).
\newblock \doi{10.1088/1742-6596/1468/1/012118}

\bibitem{Azzolini:analysis:2018}
O.~Azzolini, et~al., Eur. Phys. J. C \textbf{78}(9), 734 (2018).
\newblock \doi{10.1140/epjc/s10052-018-6202-5}

\bibitem{Andreotti_2011}
E.~Andreotti, et~al., Astropart. Phys. \textbf{34}(11), 822 (2011).
\newblock \doi{10.1016/j.astropartphys.2011.02.002}

\bibitem{Gatti:1986cw}
E.~Gatti, P.F. Manfredi, Riv. Nuovo Cimento \textbf{9}, 1 (1986).
\newblock \doi{10.1007/BF02822156}

\bibitem{Radeka:1966}
V.~Radeka, N.~Karlovac, Nucl. Instrum. Meth. A \textbf{52}, 86 (1967).
\newblock \doi{10.1016/0029-554X(67)90561-7}

\bibitem{clymene:2018}
G.~Buse, et~al., Nucl. Instrum. Meth. A \textbf{891}, 87 (2018).
\newblock \doi{10.1016/j.nima.2018.02.101}

\bibitem{Arnaboldi2011797}
C.~Arnaboldi, et~al., Astropart. Phys. \textbf{34}(11), 797  (2011).
\newblock \doi{10.1016/j.astropartphys.2011.02.006}

\end{thebibliography}

\end{document}